\documentclass[aps,pra,preprint]{revtex4-1}

\usepackage[version=3]{mhchem} % Formula subscripts using \ce{}
\usepackage[T1]{fontenc}       % Use modern font encodings

\usepackage[usenames,dvipsnames]{color}
\usepackage{graphicx}
\usepackage{dcolumn}
\usepackage{bm}
\usepackage{amsmath}
\usepackage{amsfonts}
\usepackage{psfrag}
\usepackage[bookmarks,colorlinks=false]{hyperref}
\usepackage{mathtools}
\usepackage{accents}

\usepackage[mathcal]{euscript}

\begin{document}


\title{Theory of nonlinear polaritonics: $\chi^{(2)}$ scattering on a $\beta$-SiC surface}

\author{Christopher R. \surname{Gubbin}}
\author{Simone \surname{De Liberato}}

\affiliation{School of Physics and Astronomy, University of Southampton, Southampton, SO17 1BJ, United Kingdom}

\begin{abstract}
In this article we provide a practical prescription to harness the rigorous microscopic, quantum-level descriptions of light-matter systems provided by Hopfield diagonalisation for quantum description of nonlinear scattering. A general frame to describe the practically important second-order optical nonlinearities which underpin sum and difference frequency generation is developed for arbitrarily inhomogeneous dielectric environments. Specific attention is then focussed on planar systems with optical nonlinearity mediated by a polar dielectric $\beta$-SiC halfspace. In this system we calculate the rate of second harmonic generation and the result compared to recent experimental measurements. Furthermore the rate of difference frequency generation of subdiffraction surface phonon polaritons on the $\beta$-SiC halfspace by two plane waves is calculated. The developed theory is easily integrated with commercial finite element solvers, opening the way for calculation of second-order nonlinear scattering coefficients in complex geometries which lack analytical linear solutions.
\end{abstract}

\maketitle

\section{Introduction}
At the quantum level light-matter interactions are conveniently described in a microscopic Hopfield picture, taking matter degrees of freedom as bosonic fields coupled to the photons. The system Hamiltonian is then diagonalised utilising coupled light-matter quasiparticles, termed polaritons \cite{Hopfield58}. This prescription can be followed for quantization of fields in homogeneous, linear dielectrics in the absence of \cite{Huttner91} and including \cite{Huttner92} losses. Hopfield approaches are particularly advantageous as frames to describe nonlinear processes, which can be understood as scattering cross-sections for the polaritons \cite{Carusotto13,Ciuti00,DeLiberato09,DeLiberato13,Barachati15,Savenko11,Kavokin10,Kavokin12,Liew13}.\\ 
Exploiting recent advances in photonics, demonstrated archetypically by the plasmons extant at metal-dielectric interfaces formed by hybridisation of photons with coherent oscillations of the electron gas, energy can be confined in deep subdiffraction volumes. This allows for downscaling of photonic devices to the nanoscale \cite{Gramotnev10,Kim14}  and enables measurable nonlinear scattering at diminished pump threshold. 
Plasmonic nonlinear devices often exploit intrinsic metal nonlinearities, arising from free carriers or band to band transitions and first demonstrated for harmonic generation mediated by Kretschmann-coupled surface plasmons of a sub-wavelength silver film \cite{Tsang96}. Metal-mediated plasmon generation by four-wave mixing \cite{Palomba08} and frequency difference generation \cite{Constant16} have also been demonstrated, however the centrosymmetry of plasmonic metals precludes second order nonlinear effects in the bulk and limits studies to symmetry-breaking interfaces \cite{Butet15}. Instead, theoretical studies in nonlinear plasmonics focus toward third-order nonlinear processes such as self-phase modulation \cite{DeLeon14} or other Kerr nonlinearities \cite{Huang09}.\\
Interfaces between polar and non-polar dielectrics also permit deep subdiffraction localisation by hybridising photons with coherent oscillations of the polar crystal atomic lattice, in surface phonon polaritons. Supported in the Reststrahlen region, between transverse and longitudinal optic phonon frequencies which ordinarily occupy the midinfrared spectral region, these modes absent themselves from the Ohmic losses inherent in plasmonic systems and render achievable quality factors an order of magnitude larger. Moreover, interplay between extreme energy localisation in user-defined deep sub-wavelength resonators \cite{Caldwell13,Gubbin16b} and bulk second-order nonlinearity in polar crystals make phonon polaritons an excellent testbed for midinfrared nonlinear optics \cite{Paarmann16,Razdolski16}.\\
These inhomogeneous photonic systems require polaritonic descriptions which also account for real-space variation in the dielectric environment, a necessity recently demonstrated in description of the healing length in polaritonic condensates \cite{Elistratov16}. Following initial efforts \cite{Suttorp04,Todorov14}, a full polaritonic description was recently developed for non-magnetic, arbitrarily inhomogeneous, linearly polarisable media \cite{Gubbin16c}, opening the path for quantum descriptions of nonlinear processes in sub-wavelength devices. In this Paper we embark on development of a quantum, microscopic description of second-order nonlinear scattering in inhomogeneous phonon-polariton systems. In Sec. II, after having sketched the main points of the theory of Hopfield diagonalisation in linear inhomogeneous media previously reported \cite{Gubbin16c}, we specialise our discussion to the $\beta$-SiC/vacuum interface which has been the subject of contemporary study \cite{Paarmann16}, deriving its relevant polaritonic modes. In Sec. III we develop the theory of polaritonic $\chi^{(2)}$ scattering, exploiting data regarding plane-wave scattering from recent second harmonic generation experiments \cite{Paarmann16} to both validate our theory and determining the phenomenological couplings that parameterise it. Exploiting those coefficients, that are in good agreement with data from {\it ab initio} simulations present in the literature \cite{Vanderbilt86}, we will then be able to investigate nonlinear processes involving subdiffraction phonon polariton modes. 

\section{Linear Regime}
\subsection{General theory}
Following a prior treatment \cite{Gubbin16c}, we describe a non-magnetic, inhomogeneous light-matter system, characterised by a single dipolar resonance with Hamiltonian
\begin{align}
\label{eqn:PZWH}
\hat{\mathcal{H}}_0&=\int \mathrm{d}\mathbf{r}  \left[\frac{\hat{\mathrm{D}}^2}{2\epsilon_0}+\frac{\mu_0 \hat{\mathrm{H}}^2}{2}+\frac{\hat{\mathrm{P}}^2}{2{\rho}}
+\frac{{\rho}{\omega}_{\mathrm{LO}}^2\hat{\mathrm{X}}^2}{2}-\frac{\kappa}{\epsilon_0} \hat{\mathbf{X}}\cdot \hat{\mathbf{D}}\right],
\end{align}
where $\hat{\mathbf{D}}$ $(\hat{\mathbf{H}})$ is the electric displacement (magnetic) field operator and $\hat{\mathbf{X}}$ $(\hat{\mathbf{P}})$ is the material displacement (momentum) field operator. The system is fully parameterised by spatially dependant density $\rho$, longitudinal frequency $\omega_{\mathrm{L}}$, and macroscopic dipole moment $\kappa$. In the spirit of the Hopfield prescription, the Hamiltonian in Eq.~\ref{eqn:PZWH} can be diagonalised in terms of linear superpositions of light and matter conjugate operators through the polariton operator
\begin{equation}
\label{eqn:Kexp}
\hat{\mathcal{K}}_{n}=\int \mathrm{d}\mathbf{r} \left[\boldsymbol{\alpha}_{n}\cdot \hat{\mathbf{D}} +\boldsymbol{\beta}_{n}\cdot \hat{\mathbf{H}} +\boldsymbol{\gamma}_{n}\cdot \hat{\mathbf{P}}+\boldsymbol{\eta}_{n}\cdot \hat{\mathbf{X}}\right],
\end{equation}
where the Greek symbols are the space-dependent Hopfield coefficients \cite{Gubbin16c} and $n$ indexes the different positive-frequency normal modes. Such operators obey the Heisenberg equation
\begin{equation}
\label{eqn:eigenvalues}
\left[ \hat{\mathcal{K}}_n, \hat{\mathcal{H}}_0\right] = \hbar\omega_n \hat{\mathcal{K}}_n,
\end{equation}
which are equivalent to Maxwell equations over the Hopfield coefficients in an inhomogeneous dielectric. Together with the requirement that the polaritonic operators obey bosonic commutation relations
\begin{align}
\label{eqn:BosonComm}
	\left[\hat{\mathcal{K}}_{{m}}, \hat{\mathcal{K}}_{{n}}^{\dagger}\right] &=\delta_{{m,n}},
\end{align}
this allows us to solve the system and to recover the properly normalised light and matter fields as linear superpositions of polaritonic modes weighted by the wavefunctions $\boldsymbol{\alpha}_n$
\begin{align}
	\label{eqn:ModeExpE}
	\hat{\mathbf{E}}&= \hbar \sum_{n} \omega_{n} \left[ \bar{\boldsymbol{\alpha}}_{n} \hat{\mathcal{K}}_{n} + \boldsymbol{\alpha}_{n} \hat{\mathcal{K}}_{n}^{\dag}\right],\\
	\hat{\mathbf{X}} &= \hbar \sum_{n} \frac{\kappa \omega_{n}}{\rho\left(\omega_{\text{T}}^2 - \omega_{n}^2\right)} \left[ \bar{\boldsymbol{\alpha}}_{n} \hat{\mathcal{K}}_{n} + \boldsymbol{\alpha}_{n} \hat{\mathcal{K}}_{n}^{\dag}\right]\nonumber,
\end{align}
 where the transverse resonance of the matter is 
 \begin{equation}
\omega_{\text{T}}^2 = \omega_{\text{L}}^2 - \frac{\kappa^2}{\rho \epsilon_0}. 
\end{equation}

\subsection{Application to surface phonon polaritons}
The theory sketched above can be naturally applied to the case of an halfspace filled with a polar dielectric, sketched in Figure \ref{fig:Figure1}a, identifying $\omega_{\text{L}}$ and $\omega_{\text{T}}$ with longitudinal and transverse optical phonon frequencies $\omega_{\text{LO}}$ and $\omega_{\text{TO}}$.  Developing Eq.~\ref{eqn:eigenvalues} this would lead to the solve Maxwell equations with the  inhomogeneous dielectric function
\begin{align}
\label{eqn:epsvd}
\nonumber
	\epsilon\left(\omega,z>0\right) &= \epsilon_1\left(\omega\right) =  1,\\
	\epsilon\left(\omega,z<0\right) &=  \epsilon_2\left(\omega\right) =  \epsilon_{\infty}  \frac{{\omega}_{\mathrm{LO}}^2- \omega^2}{\omega_{\mathrm{TO}}^2 -\omega^2},
\end{align}
where we labelled the mediums $1$ and $2$ and include off-resonance effects through the phenomenological high frequency dielectric constant $\epsilon_{\infty}$. These effects could be formally included by considering an additional matter resonance with longitudinal frequency far above the Reststrahlen band.
In this geometry the eigenproblem in Eq.~\ref{eqn:eigenvalues} has been solved \cite{Gubbin16c}, giving rise to seven classes of solutions. Two (TMv and TEv) describe photons incident upon the surface coming from the vacuum and their respective reflected and transmitted waves, with either Transverse Magnetic or Transverse Electric polarizations. Four (TMl, TEl, TMu, TEu) describe excitations incident upon the surface from the dielectric side and their respective reflected and transmitted waves, indexed by both the polarization and by the polaritonic branch, either lower or upper, they belong to. An example of each group is shown in Figure \ref{fig:Figure1}a.
The last one (S) is instead a surface bound evanescent solution describing the branch of surface phonon polaritons. The mode index $n$ will thus consist of the solution class and the relevant wavevector: in-plane for the evanescent solutions and both in- and out-of-plane in the medium of origin for the propagative ones.
The solution of Eq.~\ref{eqn:eigenvalues} leads to the Helmholtz equation, linking for each solution the in-plane wavevector $k_{\parallel}$ and the, generally complex, out-of-plane components $k_{i,z}$ in each halfspace $i$ by
\begin{equation}
	\epsilon_{i}\left(\omega\right)  k_0^2 = k_{\parallel}^2 + k_{i,z}^2,\label{eqn:HelmholtzSca}
\end{equation} 
where $k_0=\tfrac{\omega}{c}$ is the vacuum wavelength.
As an example in the case of propagative, TM polarised radiation incident from $z>0$ (TMv) with in- and out-of-plane wavevectors $\mathbf{k}_{\parallel}$ and $k_{1,z}$, the solution of Eq.~\ref{eqn:eigenvalues} is 
\begin{align}
\label{eqn:bilayer}
	\boldsymbol{\alpha}_{1,\mathbf{k}}^{\text{TMv}} \left(\mathbf{r}_{\parallel},z>0\right) &= N_{\mathbf{k}}^{\text{TMv}} \left[\mathbf{p}_{1 -} e^{- i k_{1,z} z} + \mathbf{p}_{1 +} r_{12}^{\text{TM}} e^{i k_{1,z} z}\right] e^{i \mathbf{k_{\parallel}}\cdot \mathbf{r}_{\parallel}},\nonumber\\
	\boldsymbol{\alpha}_{2,\mathbf{k}}^{\text{TMv}} \left(\mathbf{r}_{\parallel},z<0\right)&= N_{\mathbf{k}}^{\text{TMv}}  t_{12}^{\text{TM}} \mathbf{p}_{2 -} e^{- i k_{2,z} z} e^{i \mathbf{k_{\parallel}}\cdot \mathbf{r}_{\parallel}},
\end{align}
where $N_{\mathbf{k}}^{\text{TMv}}$ is a normalization constant to be determined by the modal orthonormality condition in Eq.~\ref{eqn:BosonComm}, $t_{12}^{\text{TM}}\; (r_{12}^{\text{TM}})$ are the Fresnel transmission (reflection) coefficients for a TM polarised beam propagating from medium 1 to medium 2 and the unit vectors for TM polarisation are given by
\begin{equation}
	\mathbf{p}_{i \pm} = \frac{k_{\parallel} {\mathbf{e}}_z \mp k_{i,z} {\mathbf{e}}_{\parallel}}{k_0 \sqrt{\epsilon_i}},
\end{equation}
where the subscript $i$ indicates the medium of propagation and the index $\pm$ the direction. Here $(\mathbf{e}_x,\mathbf{e}_y,\mathbf{e}_z)$ form a right-handed orthonormal basis and
$\mathbf{e}_{\parallel}$ is the in-plane unit vector parallel to $\mathbf{k}_{\parallel}$. The TE polarised solutions can be found by simply replacing the ${\mathbf{p}}_{i \pm}$ with the plane unit vectors perpendicular to $\mathbf{k}_{\parallel}$ and the Fresnel coefficients with their TE polarised analogues. For real incident wavevectors the normalisation constant can be calculated as
\begin{equation}
	N_{\mathbf{k}}^{\text{TMv}}= \sqrt{\frac{1}{\hbar \epsilon_0 \omega \mathcal{V}}},
\end{equation}
where $\mathcal{V}$ is the volume of quantization.
The evanescent, surface bound modes (S) shown in Figure \ref{fig:Figure2} with in-plane wavevector $\mathbf{k}_{\parallel}$ and dispersion obeying
\begin{equation}
\label{eqn:Sdisp}
k_{\parallel}=\frac{\omega}{c}\sqrt{\frac{\epsilon_{2}(\omega)}{\epsilon_{2}(\omega)+1}},
\end{equation}
have instead the form
\begin{align}
\label{eqn:bilayersphp}
	\boldsymbol{\alpha}_{1,\mathbf{k}_{\parallel}}^{\mathrm{S}} \left(\mathbf{r}_{\parallel},z>0\right) &= N_{\mathbf{k}_{\parallel}}^{\text{S}} \mathbf{p}_{1 +} e^{i k_{1,z} z} e^{i \mathbf{k_{\parallel}}\cdot \mathbf{r}_{\parallel}},\\
	\nonumber
	\boldsymbol{\alpha}_{2,\mathbf{k}_{\parallel}}^{\mathrm{S}} \left(\mathbf{r}_{\parallel},z<0\right) &= N_{\mathbf{k}_{\parallel}}^{\text{S}} \frac{\sqrt{\epsilon_{2}\left(\omega\right)} k_{1,z}}{k_{2,z}}  \mathbf{p}_{2 -} e^{- i k_{2,z} z} e^{i \mathbf{k_{\parallel}}\cdot \mathbf{r}_{\parallel}},
\end{align}
with the normalisation 
\begin{align}
	N_{\mathbf{k}_{\parallel}}^{\text{S}}&= \frac{k_{0}}{k_{\parallel}} \sqrt{\frac{\epsilon_{2}\left(\omega\right)}{\hbar \epsilon_0 \omega \mathcal{A} \left[\frac{\epsilon_{2}\left(\omega\right) - 1}{2 \lvert k_{1,z} \rvert}  + \nu\left(\omega\right)\frac{1 -  \epsilon_{2}\left(\omega\right)}{2 \lvert k_{2,z} \rvert}\right]}} \nonumber\\
	&= \frac{k_{0}}{k_{\parallel}} \sqrt{\frac{\epsilon_{2}\left(\omega\right)}{\hbar \epsilon_0 \omega \mathcal{V}_{\mathrm{eff}}\left(\omega\right)}},
\end{align}
where $\mathcal{A}$ is the quantization area and $\nu$ is the ratio of the group and phase velocities in the dielectric halfspace \cite{Gubbin16c}. We can therefore write the quantized field in the dielectric halfspace in the simple form
\begin{equation}
	\boldsymbol{\alpha}_{2,\mathbf{k}_{\parallel}}^{\mathrm{S}} \left(\mathbf{r}_{\parallel},z<0\right) =\frac{e^{i \kappa_{2,z} z} e^{i \mathbf{k_{\parallel}}\cdot \mathbf{r}_{\parallel}}}{\sqrt{\hbar \epsilon_0 \omega \mathcal{V}_{\mathrm{eff}}\left(\omega\right)}}  \left(\mathbf{e}_{\parallel} - \frac{k_{\parallel}}{\kappa_{2,z}} {\mathbf{e}}_{z} \right),
\end{equation}
with $\kappa_{2,z}=i k_{2,z}$ real and $\mathcal{V}_{\mathrm{eff}}$ can be interpreted as a dispersive effective mode volume for the evanescent mode \cite{Archambault10}.

\section{Results and discussion}

\subsection{Nonlinear Quantum Theory}
\subsubsection{General Theory}
The quadratic treatment implicit in Eq.~\ref{eqn:PZWH} amounts to the lowest order term in the Taylor expansion of the full Hamiltonian. The most general $N^{\text{th}}$ order neglected term in the case of a local interaction will consist of the product of light or matter operators
\begin{align}
\label{eqn:HNL}
\mathcal{H}_{\mathrm{NL}}^{(N)}&=\int \mathrm{d}\mathbf{r}\, \sum_{t_1\cdots t_N}\sum_{j_1\cdots j_N=1}^3\Phi^{j_1\cdots j_{N}}_{t_1\cdots t_{N}}\prod_{{l}=1}^{N} \hat{\mathcal{O}}^{j_l}_{t_l},
\end{align}
where the spatial coordinates are indexed by $j$s and the $t$s index the different light and matter operators, in our case $\hat{\boldsymbol{\mathcal{O}}}_t\in\left[\hat{\mathbf{D}},\hat{\mathbf{H}},\hat{\mathbf{X}},\hat{\mathbf{P}}\right]$.
Using the expression of the field operators in Eq.~\ref{eqn:ModeExpE} and their equivalents for the conjugate fields, we can rewrite Eq.~\ref{eqn:HNL}
in the form of a $N^{\text{th}}$ order polaritonic scattering term 
\begin{align}
\mathcal{H}_{\mathrm{NL}}^{(N)}&=\sum_{n_1\cdots n_N} 
\Xi^{n_1\cdots n_N} \prod_{l=1}^N \mathcal{K}_{n_l},
\end{align}
where the scattering tensor $\Xi$ can be found by carrying out integrals over the relevant spatial variables of the usual nonlinear tensor $\Phi$.

In the following we will aim to describe $\chi^{(2)}$ ($N=3$) effects in inhomogeneous $\beta$-SiC. As the interaction is mediated by electric field coupling to charges position, there are only three nonzero field combinations in Eq.~\ref{eqn:HNL}: the purely mechanical nonlinearity $\mathbf{\hat{X}\hat{X}\hat{X}}$, the electrical one $\mathbf{\hat{X}\hat{X}\hat{E}}$, and the Raman scattering $\mathbf{\hat{X}\hat{E}\hat{E}}$. Due to the crystal lattice symmetry, in each case the nonlinear tensor is characterised by the single parameter  coupling three orthogonal field components \cite{Roman06}, leaving us with three independent coupling coefficients to be fixed.\\
Inserting Eq.~\ref{eqn:ModeExpE} inside of Eq.~\ref{eqn:HNL} and introducing the second order susceptibility
\begin{align}
\label{eqn:NLSus}
	 \chi^{(2)}\left(\omega_1,\omega_2,\omega_3\right) &= G_1 \left[D\left(\omega_1\right) +D\left(\omega_2\right) + D\left(\omega_3\right)\right] \\ \nonumber & \quad + G_2 \left[D\left(\omega_1\right) D\left(\omega_2\right) + D\left(\omega_1\right) D\left(\omega_3\right) + D\left(\omega_2\right) D\left(\omega_3\right)\right]  \\ \nonumber & \quad + G_3 D\left(\omega_1\right) D\left(\omega_2\right)D\left(\omega_3\right),
\end{align}
with
\begin{equation}
	D\left(\omega\right) = \frac{\omega_{\text{TO}}^2}{\omega_{\text{TO}}^2 - \omega^2},
\end{equation}
the interaction Hamiltonian can be put in the usual nonlinear optics form
\begin{equation}
	\mathcal{H}^{(2)}_{\text{NL}}= \int \mathrm{d}\mathbf{r} \sum_{\mathbf{n,j_n}}
	\epsilon_0 \chi^{(2)}\left(\omega_{n_1}, \omega_{n_2},\omega_{n_3}\right)
	\hat{\mathrm{E}}_{n_1}^{j_{n_1}}\hat{\mathrm{E}}_{n_2}^{j_{n_2}}\hat{\mathrm{E}}_{n_3}^{j_{n_3}},
	\label{eqn:TNLH}
\end{equation}
where from Eq.~\ref{eqn:ModeExpE}
\begin{equation}
\hat{\mathrm{E}}^j_n=\hbar\omega_n \left[ \bar{\alpha}^j_{n} \hat{\mathcal{K}}_{n} + \alpha^j_{n} \hat{\mathcal{K}}_{n}^{\dag}\right],
\end{equation}
is the Cartesian component $j$ of the electric field in the polariton mode $n$, and the sum is over all the triplets of polariton modes with different Cartesian coordinates. 
Introducing the Born effective charge per primitive cell $Z$ \cite{Lee03}, primitive cell reduced mass $M$, and primitive cell volume $v$, we can relate the three macroscopic dimensionless coupling parameters in Eq.~\ref{eqn:NLSus} to the microscopic parameters usually derived in lattice dynamics simulations as 
\begin{align}
	G_1 &= \frac{\alpha_{TO}}{2 v} \left(\frac{Z}{M \omega_{\text{TO}}^2}\right),\\
	G_2 &= \frac{\mu^{(2)}}{2 v} \left(\frac{Z}{M \omega_{\text{TO}}^2}\right)^2,\\
	G_3 &= \frac{\phi^{(3)}}{2 v} \left(\frac{Z}{M \omega_{\text{TO}}^2}\right)^3,	\label{eqn:G3}
\end{align}
where $\alpha_{TO}$  is the Raman polarisability per primitive cell, $\mu^{(2)}$ is the second order dipole moment per primitive cell, and $\phi^{(3)}$ the third order lattice potential \cite{Roman06, Flytzanis72}. Notice that we have ignored the contribution due to the high-frequency static second order susceptibility $\chi_{\infty}^{(2)}$, that would be negligible in the resonant regime studied.

%%%%%%%%
\begin{figure}
\includegraphics[width=0.75\textwidth]{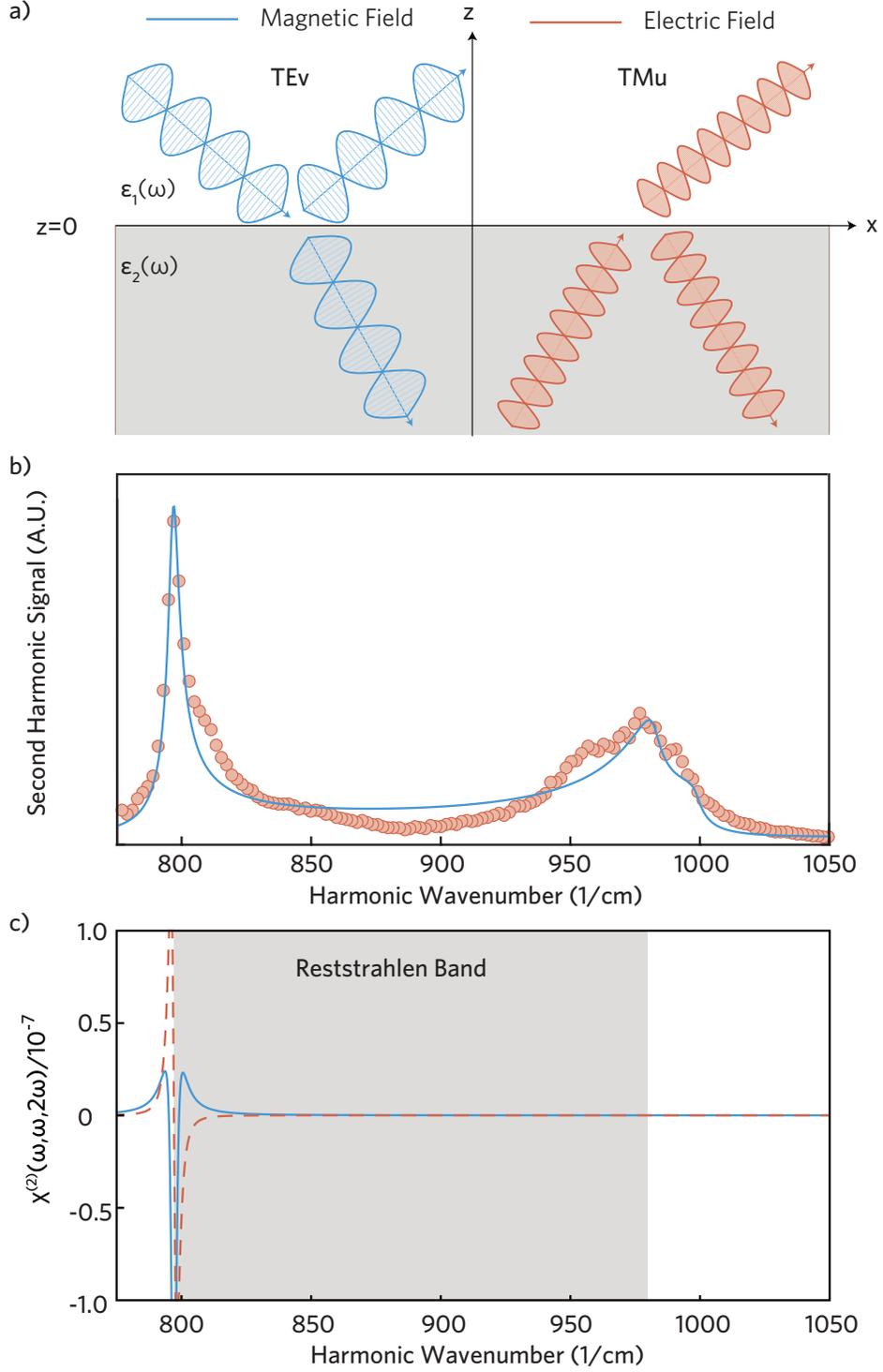}	
\caption{\label{fig:Figure1} a) Sketch of the geometry considered for the generation of a second harmonic TM signal (red) by two TE harmonic pumps (blue).  
b) Experimental data (red circles) presented by Paarmann {\it et al} \cite{Paarmann16} and theoretical fit (solid blue line) for intensity of a TM polarised second harmonic by TE polarised harmonic pumps. c) Second harmonic $\chi^{(2)}\left(\omega,\omega,2 \omega\right)$ for $\beta$-SiC. The real component is indicated by the solid blue line and the imaginary by the dashed red line. The Reststrahlen region is shaded in grey.}
\end{figure}
%%%%%%%%

\subsubsection{Second Harmonic Generation at a $\beta$-SiC/vacuum interface}
We will now apply the theory developed to the case experimentally investigated by Paarmann {\it et al} \cite{Paarmann16}: second harmonic generation at a $\beta$-SiC/vacuum interface.
We have thus to calculate the scattering tensor $\Xi$ by performing the integral in Eq.~\ref{eqn:TNLH}  using the TM polarised plane-wave eigenmodes Eq.~\ref{eqn:bilayer} or their TE polarised analogues into the nonlinear Hamiltonian and perform the space integral. As explained in Sec. II B, each modal index $n$ stands for the solution class and the relevant two- or three-dimensional wavevector in the medium of origin. For simplicity we consider the case illustrated in Figure \ref{fig:Figure1}a where the harmonic pump photons are azimuthally coplanar (all the $\mathbf{k}_{\parallel}$ are aligned).
Remembering that the nonlinear third order tensor  in zincblende crystals is characterised by a single value, coupling three orthogonally polarised fields  \cite{Vanderbilt86}, we can calculate the scattering coefficient of two TEv harmonic pumps (TE polarised incident from $z>0$, indexed by $1$ and $2$) generating a TMu second harmonic (TM polarised upper polariton mode, indexed by $3$) 
\begin{align}
 \label{eqn:Xispec}
	\Xi^{n_1,n_2,n_3} =  & \sqrt{\frac{\hbar^3\omega_{n_1} \omega_{n_2} \omega_{n_3}}{\epsilon_0 \epsilon_2\left(\omega_{n_3}\right) \mathcal{V}^3}} \chi^{(2)}\left(\omega_{n_1}, \omega_{n_2},\omega_{n_3}\right) \\&\nonumber
 \times \mathcal{T}^{\;n_1,n_2,n_3} \delta\left(\mathbf{k}_{\parallel}^{n_1} + \mathbf{k}_{\parallel}^{n_2} - \mathbf{k}_{\parallel}^{n_3}\right) \\
 &\nonumber \times \left[ \mathcal{P}\left(\frac{1}{i \Delta k_z^{n_1,n_2,n_3}}\right)  - \pi \delta\left(\Delta k_z^{n_1,n_2,n_3}\right)\right],
\end{align}
where $\mathcal{P}$ indicates the principal value and
the choice of a TMu second harmonic is fixed by the requirements of having a single rays escaping toward $z>0$ and a resonance frequency larger than 
$\omega_{\text{LO}}$.
The mismatch in the out-of-plane wavevectors in Eq.~\ref{eqn:Xispec} is given by
\begin{equation}
\Delta k_z^{n_1,n_2,n_3} = k_{2,z} \left(\mathbf{k}_{\parallel}^{n_1}, k_{1,z}^{n_1}\right) + k_{2,z}\left(\mathbf{k}_{\parallel}^{n_2}, k_{1,z}^{n_2}\right) + k_{2,z}^{n_3},
\end{equation}
and the Fresnel tensor has the form
\begin{align}
\label{L}
	\mathcal{T}^{\;n_1,n_2,n_3}&= 2 \frac{k_{\parallel}^{n_3}}{k_0^{n_3}} \left({\mathbf{e}}_{\parallel}^{n_1} \cdot {\mathbf{e}}_x\right) \left({\mathbf{e}}_{\parallel}^{n_1} \cdot {\mathbf{e}}_y\right) \\&\nonumber \quad \times L_{yy}\left(\mathbf{k}_{\parallel}^{n_1},  k_{1,z}^{n_1} \right)  L_{yy}\left(\mathbf{k}_{\parallel}^{n_2},  k_{1,z}^{n_2} \right), 
\end{align}
with
\begin{align}
	L_{yy} \left(\mathbf{k}_{\parallel}, k_{1,z} \right) &= \frac{2  k_{1,z}}{  k_{1,z} + k_{2,z} \left(\mathbf{k}_{\parallel}, k_{1,z} \right)},
\end{align}
and the non-indexed out-of-plane wavevectors and the frequencies are calculated through the Helmholtz equation in Eq.~\ref{eqn:HelmholtzSca}. 
Utilising Fermi's golden rule and summing over final states with different out-of-plane wavevectors we can now calculate the scattering rate in the TM upper polariton mode of frequency $\omega$ 
\begin{align}
\label{eqn:FGR}
	\mathcal{W} =& \frac{ \hbar N_{1}N_{2} \omega_{n_1}\omega_{n_2} }{16 \pi^4 \epsilon_0 } \frac{\mathcal{A}}{\mathcal{V}^2} \int d\omega\; \omega\rho(\omega,\lvert \mathbf{k}_{\parallel}^{n_1}+\mathbf{k}_{\parallel}^{n_2}\lvert)\\
	& \times \biggr|  \frac{\chi^{(2)}\left(\omega_{n_1}, \omega_{n_2},\omega\right)}{i \Delta k_z^{n_1,n_2,n_3}} \frac{\mathcal{T}^{\;n_1,n_2,n_3}}{\sqrt{\epsilon_2\left(\omega_{n_3}\right)}}\nonumber 
	 \biggr|^2  \delta(\omega_{n_1}+\omega_{n_2}-\omega), \nonumber
\end{align}
where $N_{j}$  is the photon number in the pump beam $j$ and
\begin{equation}
	\rho\left(\omega, {k}_{\parallel}\right)  = \frac{\partial k_{2,z}}{\partial \omega} = \frac{\sqrt{{k_{2,z}}^2+k_{\parallel}^2}}{k_{2,z}} \frac{1}{\mathrm{v_g}\left(\omega\right)},
\end{equation}
with $\mathrm{v_g}(\omega)$ the group velocity. Note that in Eq.~\ref{eqn:FGR} the term proportional to $ \delta\left(\Delta k_z^{n_1,n_2,n_3}\right)$ disappears because for the second harmonic generation we wish to consider $\Delta k_z^{n_1,n_2,n_3}=0$ cannot be satisfied together with the conservation of energy enforced by  $\delta(\omega_{n_1}+\omega_{n_2}-\omega)$.
Specialising Eq.~\ref{eqn:FGR} to the case of second harmonic generation with harmonic pump frequency $\omega_0$,  and defining the energy density for each pump $\mathcal{F}_j = \hbar \omega_0 N_{j}/ \mathcal{V}$, we finally obtain the emission rate per unit of surface
\begin{align}
	\frac{\mathcal{W}}{\mathcal{A}} &= \frac{\mathcal{F}_1\mathcal{F}_2 \omega_0}{8\pi^4 \hbar \epsilon_0 } \rho(2\omega_0) \biggr|  \frac{\chi^{(2)}\left(\omega_0, \omega_0,2\omega_0\right)}{i \Delta k_z^{n_1,n_2,n_3}} \frac{\mathcal{T}^{\;n_1,n_2,n_3}}{\sqrt{\epsilon_2\left(2\omega_0\right)}} \biggr|^2.
	\label{eqn:SHG}
\end{align}
In order to quantitatively fit the experimental results \cite{Paarmann16} we now need to introduce the effect of losses into Eq.~\ref{eqn:SHG}. While we could have used from the beginning the lossy version of the theory \cite{Gubbin16c}, this would have been needlessly cumbersome given the low losses at the relevant frequencies in SiC, that mainly only result in a broadening of the resonances. 
We can thus simply generalise Eq.~\ref{eqn:SHG} to the case of weak loss by considering a material damping $\gamma$ in the dielectric function in Eq.~\ref{eqn:epsvd}
\begin{align}
\label{eqn:epsdiss}
\tilde{\epsilon}_2\left(\omega\right)=  \epsilon_{\infty}  \frac{{\omega}_{\mathrm{LO}}^2- \omega^2}{\omega_{\mathrm{TO}}^2 -\omega^2-i\gamma\omega},
\end{align}
and keeping into account the presence of the complex poles in the dispersive functions $D\left(\omega\right)$ comprising $\chi^{(2)}$ in Eq.~\ref{eqn:NLSus} 
\begin{equation}
\label{eqn:Ddiss}
	\tilde{D} \left(\omega\right) = \frac{\omega_{\text{TO}}^2}{\omega_{\text{TO}}^2 - \omega^2 - i \omega \gamma}.
\end{equation}
Using the lossy version of Eq.~\ref{eqn:SHG} we can fit to the experimental data \cite{Paarmann16}, where a free electron laser is used to measure the second harmonics generated in reflection at a $\beta$-SiC surface. 
Notice that form the the form of the normal modes in Eq.~\ref{eqn:bilayer}, and as clearly shown in Figure \ref{fig:Figure1}a, the emitted TMu modes contain both a part propagating inside the dielectric and one part radiating in vacuum. As the experiment collects and measures only the latter part, in order to correctly fit the experiments we need to multiply the emission rate obtained from Eq.~\ref{eqn:SHG} for the Fresnel power transmittance coefficient for the upward travelling outgoing beam $\lvert t_{21}^{\mathrm{TM}} \rvert^2$.
In Figure \ref{fig:Figure1}b we show the theoretical result from Eq.~\ref{eqn:SHG} (blue solid line) compared to the experimental data, kindly made available by Paarmann {\it et al} \cite{Paarmann16} (red dots). Our theory correctly replicates the position of the two main resonances seen in the experimental data without adjustable parameters.
The peak near $\omega_{\mathrm{LO}}$ results from a resonance in the pump Fresnel tensor, while the one at $\omega_{\mathrm{TO}}$ is a result of strong resonance of the second order susceptibility, plot for a second harmonic process in  Figure \ref{fig:Figure1}c, due to the pole in Eq.~\ref{eqn:Ddiss}.   
 
In order to quantitatively reproduce also the peak intensities we used the coupling coefficients $G_1, \;G_2,\;G_3$ introduced in in Eq.~\ref{eqn:NLSus} as fitting parameters, yielding a good quantitative agreement and thus fixing the ratios of the different coupling constants.
Combining these results with measurements of the Faust-Henry coefficient $G_1/ \chi_{\infty}^{(2)}$ which describes the Raman anharmonicity \cite{Yugami86} and of the unclampled-ion, strain free, electro-optic susceptibility \cite{Tang91}
\begin{equation}
	\chi^{(2)}_{\mathrm{eo}} = \chi_{\infty}^{(2)}+G_1,
\end{equation}
it is therefore possible to calculate the absolute magnitude of the coupling parameters $G_1, \;G_2,\;G_3$ which fully define the second-order nonlinear susceptibility in the neighbourhood of the lattice optical phonon resonances. The third-order lattice potential $\phi^{(3)}$ is thus easily calculable from the fit by Eq.~\ref{eqn:G3} and it could be also independently calculated by {\it ab initio} methods. While such {\it ab initio} simulations for $\beta$-SiC are not present in the literature, and they are beyond the scope of the present paper, an estimate can be derived by the frozen phonon calculations of Vanderbilt {\it et al.} for the third-order lattice potentials of monoatomic crystals in diamond structure \cite{Vanderbilt86}.
In such a work the authors note that the zone-center nonlinear coefficients rescaled on the bond length $a$ and spring constant $g$
\begin{equation}
	\tilde{\phi}^{(3)} = \frac{a}{g} \phi^{(3)},
\end{equation}
present a remarkable universality for materials as different as C, Si, and Ge, for which $\tilde{\phi}^{(3)} \approx - 4$. Extrapolating this universality to be at least partially valid for $\beta$-SiC, that shares both crystal structure and atomic constituents with them, we obtain
a qualitative agreement with our result $\tilde{\phi}^{(3)} = -1.35$. 

%%%
\subsection{Difference Generation of SPhPs}
\begin{figure}
\includegraphics[width=0.75\textwidth]{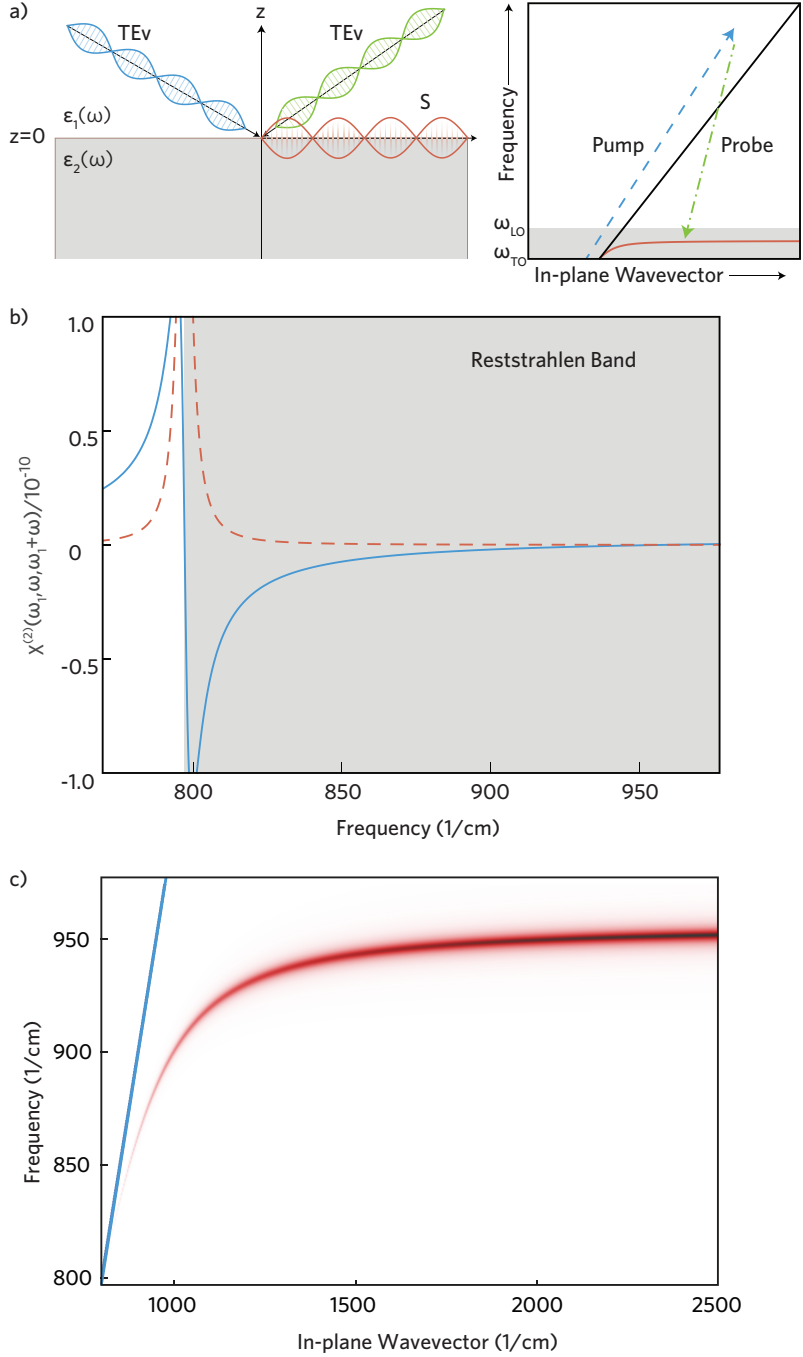}
\caption{\label{fig:Figure2} a) Schematic of the generation of a low frequency evanescent wave (red) by high frequency pump (blue) and probe (green) beams. For the sake of clarity only the incident waves in the vacuum halfspace are shown. b) The non-linear susceptibility for the process illustrated in a), as a function of signal beam frequency. c) Normalised spectrum of the difference frequency generation rate as a function of the emitted surface mode wavevector obtained by varying the probe frequency and inclination with a fixed pump, as described in the main text. The blue line represents the light line in vacuum.}	
\end{figure}
%%%
Having fixed material parameters through reproducing experimental results we are now in a position to make predictions about the rates for novel non-linear processes. The excitation of subdiffraction surface modes by free space radiation was first proposed by Novotny {\it et al.}, utilising a four-wave mixing procedure to excite the plasmons of a gold halfspace \cite{Palomba08}. A four-wave mixing procedure is necessary to conserve energy on excitation of modes in the visible spectral region utilising common sources emitting in the UV-visible spectral region. Subdiffraction modes in the midinfrared spectral region however can be generated by three-body scattering of pump beams operating in the visible spectral region. This was recently demonstrated to excite the low-energy plasmons of monolayer graphene from 5 to 50THz \cite{Constant16}. 
Utilising the quantization for the evanescent S modes  Eq.~\ref{eqn:bilayersphp} we can describe to the process illustrated in Figure\ref{fig:Figure2}a, describing difference generation of surface phonon polaritons. At the quantum  level the process can be described by decay of a photon from the pump into a surface mode accompanied by stimulated emission of a photon into the probe. Considering for definiteness the matrix elements for such a process involving TE polarised pump (1) and probe (2) beams with a surface evanescent signal (3) we find
\begin{align}
	\Xi^{n_1,n_2,n_3} =&  \sqrt{\frac{\hbar^3 \omega_{n_1} \omega_{n_2} \omega_{n_3}}{\epsilon_0 \epsilon_2\left(\omega_{n_3}\right) \mathcal{V}_{\mathrm{eff}}\left(\omega_{n_3}\right)\mathcal{V}^2 }}  \\
	 &\times \frac{\chi^{(2)}\left(\omega_{n_1}, \omega_{n_2},\omega_{n_3}\right)}{i \Delta k_{z}^{n_1,n_2,n_3}} \mathcal{T}^{\;n_1,n_2,n_3} \delta\left(\mathbf{k}_{\parallel}^{n_1} - \mathbf{k}_{\parallel}^{n_2} - \mathbf{k}_{\parallel}^{n_3}\right)\nonumber,
\end{align}
where the Fresnel tensor now reads
\begin{align}  
	\mathcal{T}^{\;n_1,n_2,n_3} = &2 \left({\mathbf{e}}_{\parallel}^{n_1} \cdot {\mathbf{e}}_x\right) \left({\mathbf{e}}_{\parallel}^{n_1} \cdot {\mathbf{e}}_y\right) \\ &\times L_{yy}\left(\mathbf{k}_{\parallel}^{n_1}, k_{1,z}^{n_1} \right)  L_{yy}\left( \mathbf{k}_{\parallel}^{n_2}, k_{1,z}^{n_2} \right),\nonumber
\end{align}
and the relevant second order non-linear susceptibility of the $\beta$-SiC halfspace as a function of the signal frequency is shown in Figure \ref{fig:Figure2}b. This is substantially less than the nonlinear susceptibility for second harmonic generation illustrated in Figure \ref{fig:Figure1}c, because only the signal beam is close to the $\omega_{\text{TO}}$ resonance. 
The surface mode scattering rate can be now determined by assuming Lorentzian broadening for a surface mode with in-plane wavevector $\mathbf{k}_{\parallel}$
\begin{equation}
	\rho^{\text{S}}\left(\omega,{k}_{\parallel}\right) = \frac{1}{\pi} \frac{\mathrm{Re}\left[\omega^{\mathrm{S}}_{k_{\parallel}}\right]}{\left(\omega -\mathrm{Re}\left[\omega^{\mathrm{S}}_{k_{\parallel}}\right] \right)^2 + \mathrm{Im}\left[\omega^{\mathrm{S}}_{k_{\parallel}}\right]^2},
	\label{eqn:SPhPdos}
\end{equation}
where the surface mode real and imaginary frequencies $\mathrm{Re}\left[\omega^{\mathrm{S}}_{k_{\parallel}}\right]$ and $\mathrm{Im}\left[\omega^{\mathrm{S}}_{k_{\parallel}}\right]$ are solutions of Eq.~\ref{eqn:Sdisp} with the dissipative dielectric function in Eq.~\ref{eqn:epsdiss}. 
Fixing the pump's frequency $\omega_1=17000$cm$^{-1}$ and its incidence angle to $30^{\circ}$ (thus effectively fixing the in-plane wavevector) we can now vary the probe frequency and inclination, leading to resonant emission at frequency $\omega$ in the surface mode described by the emission rate per unit surface
\begin{align}
\label{eqn:EmissionS}
	\frac{\mathcal{W}}{\mathcal{A}}& = \frac{\mathcal{F}_{1} \mathcal{F}_{2}  \omega }{8 \pi^4 \epsilon_0 \mathcal{L}_{\mathrm{eff}}\left(\omega\right)\hbar \; \mathrm{Im}\left[\omega^{\mathrm{S}}_{k_{\parallel}}\right]} \\ 
	& \quad \times \nonumber \biggr| \frac{\chi^{(2)}\left(\omega_{1}, \omega_{1}-\omega,\omega\right)}{i \Delta k_{z}^{n_1,n_2,n_3}} \frac{\mathcal{T}^{\;n_1,n_2, n_3}}{\sqrt{\epsilon_{2}\left(\omega\right) }} \biggr|^2,
\end{align}
where $\mathcal{L}_{\mathrm{eff}}=\mathcal{V}_{\mathrm{eff}}/\mathcal{A}$ is the effective confinement length along the $z$ direction.

\subsection{Second Harmonic Generation in Subdiffraction Resonators}
From Eq.~\ref{eqn:SHG} and Eq.~\ref{eqn:EmissionS} we can see that the intensity of second-order nonlinear response can be modified engineering two main parameters. First the $\chi^{(2)}$ of the material facilitating the nonlinear interaction which, for phononic nonlinearities, peaks at the transverse optical phonon frequency. Second the overlap between the participating modes, leading to the effective confinement factor in Eq.~\ref{eqn:EmissionS} when one or more of the modes are evanescent or localised.
Both of these factors can be improved by exploiting the morphologically dependant resonances of subdiffraction $\beta$-SiC resonators which have recently been investigated experimentally \cite{Caldwell13, Gubbin16a} and numerically \cite{Gubbin16b, Chen14}. These modes provide spectral tuneablility of the strong field enhancement, allowing a greater overlap of the confined mode with the resonance of the material $\chi^{(2)}$. In addition by confining the mode in three dimensions it is possible to achieve ultrasmall mode volumes up to $10^{6}$ times smaller than the wavelength limited volume in SiC, which provide strong enhancements to the local photonic density of states. Recently second harmonic generation exploiting the resonances of subdiffraction 4H and 6H-SiC resonators was demonstrated experimentally \cite{Razdolski16}. In this work the interplay between field confinement and the dispersion of the $\chi^{(2)}$ was directly observed in the form of resonant enhancement of second harmonic generation in the region of a band-folded lattice phonon branch. Our model can be readily extended to describe these systems utilising numerical simulations of the quantized system eigenmodes in the linear regime \cite{Gubbin16b}. The overlap integrals in the nonlinear Hamiltonian Eq.~\ref{eqn:HNL} can then be readily calculated.

\section{Conclusion}
In conclusion we have presented a comprehensive quantum theory of $\chi^{(2)}$ scattering in inhomogeneous light-matter systems, applying it to model recent nonlinear experiments on $\beta$-SiC surfaces. The quantitative agreement between the resonant frequencies predicted by our theory and the experimental data previously reported \cite {Paarmann16} validated our theory, while a fitting of the relative peak heights allowed us to fix the phenomenological scattering parameters. With a fully parameterised theory we were then able to make quantitative prediction of novel nonlinear effects.
Phonon polaritons, either on flat surfaces, or localised in sub-wavelength resonators, have the potential to become an important platform for midinfrared photonics and quantum polaritonics. The present paper is a first, important step in gaining a quantitative understanding of their scattering properties at the quantum level, necessary to open the door to future investigations on the possibility to realise quantum fluids of light \cite{Carusotto13} made of phonon polaritons, in a sort of midinfrared analogous of the very successful investigations of microcavity exciton-polaritons in the near-infrared region.

\section{Acknowledgements}
The authors thank Alex Paarmann for provision of data used to fit our scattering coefficients. We acknowledge support from EPSRC grant EP/M003183/1. S.D.L. is Royal Society Research Fellow.

\end{document}